\documentclass[a4paper]{article}
\usepackage{graphicx}
\usepackage{latexsym}
\usepackage{amssymb,amsfonts,amsmath}
\usepackage{epsfig,epic}
\usepackage[dvips]{color}
\title{Overlap of the Wilson loop \\ with the broken-string state}
\author{F.Gliozzi$^a$ and  A.Rago $^b$}
\newcommand{\eq}{\begin{equation}}
\newcommand{\en}{\end{equation}}
\newcommand{\ear}{\begin{eqnarray}}
\newcommand{\rae}{\end{eqnarray}}
\newcommand{\Z}{\mathbb{Z}}
\newcommand{\bra}{\langle}
\newcommand{\ket}{\rangle}
\newcommand{\um}{\frac12}
\definecolor{M_Beige}         {rgb}{0.96 , 0.96 , 0.86}

\definecolor{M_Brown}         {rgb}{0.65 , 0.16 , 0.16}

\definecolor{M_Gold}          {rgb}{1.00 , 0.84 , 0.00}

\definecolor{M_LemonChiffon}  {rgb}{1.00 , 0.98 , 0.80}

\definecolor{M_Orange}        {rgb}{1.00 , 0.60 , 0.00}

\definecolor{M_Pink}          {rgb}{1.00 , 0.75 , 0.80}

\definecolor{M_Violet}        {rgb}{0.93 , 0.51 , 0.93}
\begin{document}
%\begin{flushright}
%IFUM-793-FT
%\end{flushright}
\maketitle
\noindent
 $^a$ Dipartimento di Fisica Teorica, Universit\`a di Torino and\\ INFN,
sezione di Torino, via P. Giuria, 1, I-10125 Torino, Italy.\\
$^b$ Dipartimento di Fisica, Universit\`a di Milano and\\
INFN, Sezione di Milano, Via Celoria, 16, I-20133 Milano, Italy.
\vskip0.5cm\noindent
e-mail: gliozzi@to.infn.it, antonio.rago@mi.infn.it\\
\begin{minipage}[0cm]{\textwidth}
\vspace{-16cm}
\hfill IFUM-793-FT
\end{minipage}
\begin{abstract}
Numerical experiments on most gauge theories coupled with matter 
failed to observe string-breaking effects while measuring Wilson loops 
only.  We show that, under rather mild assumptions, the overlap of the Wilson 
loop operator with the broken-string state obeys a natural upper 
bound implying that the signal of string-breaking is in general too 
weak to be detected by the conventional updating algorithms.

In order to reduce the 
variance of the Wilson loops  in 3-D $\Z_2$ gauge Higgs model 
we use a new algorithm based on  the  L\"uscher-Weisz method  combined 
with a non-local cluster algorithm which allows to follow the decay of 
rectangular Wilson loops up to values of the order  of $10^{-24}$.  
In this way a sharp signal of string breaking is found. 

\end{abstract}

\section{Introduction}

The confining force between a pair of static sources in pure gauge theories 
is mediated by a thin flux tube, or string, joining the two sources.
When matter is added to this system the string becomes unstable at large 
separations and breaks when it reaches a certain length $R_b$ to form 
pairs of matter particles. This  breaking should produce  the screening  
of the confining force between the  sources and hence the flattening of 
the static potential.  

The lack of any sign of flattening in most systems,   
while measuring the static potential from  Wilson loops only, came as a 
surprise \cite{qcd,pt}.
 
One suggestion to arise in literature is that the Wilson loop has  a  
poor overlap with the true ground state, hence the basis of the 
operators has to be enlarged \cite{cm} in order to get a reliable
estimate of the potential. Using this multichannel method it has been 
observed the breaking of the confining string in Higgs models \cite{ks} in QCD 
\cite{milc,bdl} and in the SU(2) Yang Mills theory with adjoint sources \cite{st}.
Even if some cautionary observations have been raised about this method
\cite{kt}, one is led to conclude that the difficulty in observing  string 
breaking  with the Wilson loop seems to indicate nothing more than that it 
has a very small overlap with the broken-string state.

This fact has been directly demonstrated  in $2+1~d$ SU(2) YM theory with
 adjoint sources \cite{krde} where,
using a variance reduction algorithm allowing to detect signals down to 
$10^{-40}$, it has been clearly observed a rectangular Wilson loop 
 $W(R>R_b,T)$
 changing sharply its slope as a function of $T$ from that associated to 
the unbroken string (area-law decay) to that of the broken-string state 
(perimeter-law decay) at a distance much longer than the string breaking scale
$R_b$. Here we will undertake a similar work for the $2+1~d$ $\Z_2$ 
gauge-Higgs model  and find a similar result. With the emergence of this 
long-distance effect  a closely related question comes in: why the 
overlap of the Wilson loop with the broken-string state is so small?.

\par Before trying to answer to this question, we should mention that 
do exist gauge systems where such overlap is much bigger. For instance, 
in the $2+1~d$  $\Z_2$  gauge-Higgs  \cite{gr} and in $3+1~d$ SU(2) gauge-Higgs 
models \cite{Bertle:2001ya} it has been identified a  region of the  space 
of the coupling constants where the vacuum has a rather unusual property 
in that it has, besides the magnetic monopole condensate 
characterising the confining ``phase'', also a non-vanishing electric 
condensate, like in the dual Higgs phase. We argued \cite{gr} that in this 
region the world sheet of the Wilson loop belongs to the so called tearing 
phase \cite{kaza,gp}, characterised by the formation of holes of arbitrary 
large size, reflecting pair creation. As a consequence, larger Wilson loops 
give rise to larger holes and the vacuum expectation value follows 
the perimeter-law, signalling string breaking. Actually, measuring Wilson 
loops in such a special region of the $2+1~d~ \Z_2$ gauge Higgs model, we 
observed \cite{gr} a smooth transition from an area decay at relatively short 
distances to a perimeter decay at long distances.
\par
Hints of an appreciable overlap of large Wilson loops with the broken-string 
ground state have been also reported in 2+1 dimensional SU(2) gauge 
theory with two flavours \cite{tr} using highly anisotropic improved lattices 
 and even in 3+1 QCD \cite{det}, measuring Wilson line correlators in Coulomb 
gauge with an improved action.
\par
In this paper we consider instead isotropic lattices with standard plaquette 
action in the confined ``phase''.  Our main goal  is to understand why in 
these cases, for all coupled gauge systems studied thus far, it is so difficult 
to observe string breaking using Wilson loops only. 
\par
It was argued \cite{gp} that in these cases the   
world sheet associated to the confining string belongs to the normal phase, 
where the holes induced by dynamical 
matter have a mean size which does not depend on that of the Wilson loop: 
larger loops give rise to the formation of more holes. 
As a consequence they decay with an area-law 
as in the quenched case. This suggests we assume the following Ansatz for 
the asymptotic behaviour of large, rectangular Wilson loops 
\footnote{For sake of simplicity we  momentarily neglect the universal 
contribution  of the quantum fluctuations of the flux tube. For an improved
 Ansatz see Eq.(\ref{refan}) below.}
\eq 
 W(R,T)\simeq c_u\exp[-2\rho(R+T)-\sigma RT]+c_b\exp[-2\mu (R+T)]
\label{Ansatz} 
\en
The first term describes the typical area-law decay of  pure Yang-Mills 
theory with a string tension $\sigma$. The second term is  instead 
the contribution expected in the broken-string state; it  decays with a 
perimeter law controlled by the mass $\mu$ 
of the so called static-light meson, the lowest bound state of a static 
source and a dynamical Higgs field or quark . 
\par
In \cite{gp} it was even conjectured that the perimeter term could be zero
(no overlap). Note however that any Wilson loop of finite size receives 
contributions not only from world sheet configurations typical of the normal 
phase, but also from those of the tearing phase 
\cite{gg}, hence $c_b\not=0$. 
\par
It is easy to see \cite{krde} that  for  $R<R_b$ 
and $c_b$ small enough the first term dominates over the whole range of 
$T$  and no sign of string-breaking can be seen. 
When $R>R_b$, no 
matter how small $c_b$ is, the above Ansatz 
implies that at  long distances the broken string 
behaviour eventually prevails, since the first
 term drops off more rapidly than the second. 
We must of course also require that the latter should not dominate 
 at distances shorter than the string breaking scale $R_b$, thus it is 
obvious that $c_b$ cannot be too big. 
\par
A closer look shows (Section \ref{upper}) that the ratio $c_b/c_u$ is 
bounded from above by the following inequality
\eq
\log \frac{c_b}{c_u}\leq\,\sigma R_f\,(2R_b-R_f)~~,
\label{bound}
\en
where $R_f$ is the minimal scale of string formation. Inserting this upper
bound into (\ref{Ansatz}) one easily checks on the back of an envelope that as 
$R$ ranges from, say, $R_b$  to $2\,R_b-R_f$,
the Wilson loop $W(R,T)$ could deviate from the unbroken 
string behaviour only for $T\geq 2\,R_b-R_f$ , where its vacuum 
expectation value is in general exceedingly small 
(Section \ref{upper}). 
\par
Thus such a bound gives us  a simple explanation of the fact that it is 
so difficult to see a signal of string breaking while measuring Wilson loop 
only. It can  also be used as a guide to search in the parameter space of the
gauge theory a region where it is possible to evaluate the vacuum expectation 
value of rectangular Wilson loops with $R>R_b$ and  $T>2\,R_b-R_f$.
\par
We applied that procedure to the 2+1 dimensional $\Z_2$ gauge-Higgs model 
(Section \ref{model}). 
Although such a system  is perhaps the simplest example of a gauge 
theory coupled to matter in the fundamental representation, 
the current algorithms are not sufficiently accurate for our purpose. Thus we 
developed a new one, based on a version of the recent L\"uscher-Weisz 
variance reduction \cite{lw} for updating the gauge degrees of freedom, 
combined with a non-local cluster algorithm for the matter degrees of 
freedom ( Section \ref{algorithm}). 
\par
The  new  
 algorithm allowed us  to detect signals down to 
$10^{-24}$. We observed in this way rectangular Wilson loops 
$W(R,T\simeq 2\,R_b)$ 
 changing abruptly their slope as a function of $R$ as described by the 
Ansatz (\ref{Ansatz}).  It turns out that the breaking point nearly saturates 
 the upper bound (Section \ref{results}).

\section{The upper bound}
\label{upper}
  
Consider a gauge system composed by a gauge field coupled to whatever kind of 
matter. The confining string between a pair of static 
sources should be unstable against breaking at large $R$, where 
dynamical matter particles can materialise to bind to the static sources, 
forming a pair of bound states called static-light mesons. 

Denoting by $\mu$ the mass of the lowest bound state, the static potential 
is expected to approach the constant value
\eq
\lim_{R\to\infty}V(R)=2\,\mu~~,
\label{vinf}
\en
where
\eq
V(R)=-\lim_{T\to\infty}\frac 1T W(R,T)~~.
\en
Although Eq (\ref{vinf}) refers to an  asymptotic value, numerical 
simulations on these systems show that for all practical purposes 
we may safely assume
\eq
V(R)\simeq 2\,\mu~~,~~~~ R>R_b~~~, 
\label{vmu}
\en
while in the range $R_f\leq R \leq R_b$, where the confining string is
 formed and is stable, the potential should have the typical confining form 
dictated  by the first term of Eq. (\ref{Ansatz})
\eq
V(R)\simeq 2\rho+\sigma R~~,~~~R_f\leq R\leq R_b~~~.
\label{vconf}
\en
Combining Eq.(\ref{vmu}) and (\ref{vconf}) yields
\eq
R_b\simeq\frac{2\mu-2\rho}{\sigma}~~.
\label{rb}
\en
Notice that the mass $\mu$ and the perimeter term $\rho$ 
are not UV finite because of the additive divergent self-energy 
contributions of the static sources. However, these divergences should 
cancel in their difference, hence $R_b$ is a meaningful physical scale 
even in the continuum limit.
\par
In the range 
$R_f\leq R\leq R_b$, according to Eq.(\ref{vconf}), the first term of 
the Ansatz (\ref{Ansatz})  dominates over the second in the limit 
$T\to\infty$, hence it should also dominate  for any finite $T$, 
because 
 $V(R)$ is less than $2\,\mu$ in this range of $R$. Thus
\eq 
c_u\, e^{-2\rho(R+T)-\sigma RT}\geq c_b\, e^{-2\mu(R+T)}~,~  
R_f\leq R\leq R_b~,~
T>R_f~.
\label{Ineq}
\en
With the help of Eq.(\ref{rb}), this inequality can eventually be recast into 
 the form
\eq
\log\frac{c_b}{c_u}\leq\sigma\,\left[R_b(R+T)-RT\right]~~,
~R_f\leq R\leq R_b,~T>R_f~.
\label{ineq}
\en
Putting $R=T=R_f$ yields the sought after bound (\ref{bound}). For a 
graphical representation of the Ansatz and the subsequent bounds see
Fig. \ref{Figure:1}.
\par
In the confining string picture, the inequality (\ref{Ineq}) tells us
that a string of length $R_f\leq R\leq R_b$ is stable and then it can 
propagate for
 any interval of  $T$. A natural question comes to mind. What happens 
when we stretch the string beyond  $R_b$? clearly it becomes unstable 
against breaking. Let  ${\cal T}={\cal T}(R)$ be the amount of ``time''
it takes for a string of length $R$ to break, 
defined by the equality of the two terms of the Ansatz (\ref{Ansatz}).
Inserting there the upper bound (\ref{bound}) we can find 
a lower bound for  ${\cal T}$  as a function of $R$:
\eq
{\cal T}(R)\geq \frac{R_b\,(R-2\,R_f)+R_f^2}{R-R_b}~~,~R>R_b~.
\label{survival}
\en
${\cal T}$ represents the minimal survival time of a string of length $R$
 before breaking. In order to maximise the signal 
one must minimise the area $A(R)=R\,{\cal T}(R)$ of the Wilson loop. Looking 
at the solution $R_o$ of $\frac{d\,A(R)}{d\,R}=0$ 
 one gets after a little algebra
\eq
T_b\equiv {\cal T}(R_o)=2R_b-R_f=R_o~.
\label{tb}
\en
It represents the most favourable distance for observing string breaking in 
the case in which the upper bound were saturated; but even in this 
limit case the signal is very weak, being proportional to 
$e^{-\sigma T_b^2} $.  From a 
computational point of view it is very challenging  to reach such 
length scales in the measure of $ W(T,R) $ even in the 
simplest models. This explains why it is so difficult 
to see this effect.
\par Thus far we have neglected the effects due to quantum string 
fluctuations. A more accurate Ansatz which accounts for these  will be 
used while fitting the numerical data (see Eq.(\ref{refan})). Had we used 
such a refined form in deriving the above bounds, we would have obtained 
much more involved formulae without modifying very much their 
numerical value.
\par
In deriving the previous inequalities it was assumed that the 
confining string is created by the Wilson loop. It should be clear that 
these formulae are not applicable to strings generated by different operators.
Notice that different operators correspond  to different boundary conditions 
of the string world sheet and these imply in turn different long distance 
behaviour. For instance, in the case of a coupled gauge theory at 
finite temperature, it is quite obvious how to modify the Ansatz 
(\ref{Ansatz}) in order to describe the Polyakov line correlator 
(see {\sl e.g.} Eq.(10) of \cite{gp}). In particular the second term is 
just a constant at any fixed
 temperature; this implies that the overlap of the Polyakov line correlator 
 with the broken string state is maximal above $R_b$, in agreement with the 
fact that at finite temperature string breaking has been easily 
seen \cite{finte}.  
\section{$\Z_2$ gauge-Higgs  action and observables}
\label{model}
The action of a  2+1 dimensional $\Z_2$ gauge theory coupled to Ising  
 matter in a cubic lattice $\Lambda$ can be written as 
\eq
S(\beta_G,\beta_I)=-\beta_I\sum_{\langle ij\rangle}
\varphi_i U_{ij}\varphi_j-
\beta_G\sum_{plaq.}U_{\square}~,
\en
where both the link variable $U_{ij}\equiv U_\ell$  and the matter
field $\varphi_i$  take values $\pm1$ and
$U_{\square}=\prod_{\ell\in\square}U_\ell$.

This model is self-dual: the Kramers-Wannier transformation maps the
model into itself. Its partition function
\eq
Z(\beta_G,\beta_I)=\int[\mbox{DU]\,[D}\varphi]\,
e^{-S(\beta_G,\beta_I)}
\en
fulfils  the functional equation
\eq
Z(\beta_G,\beta_I)=(\sinh 2\beta_G \sinh 2\beta_I)^{\frac 32 N}
Z(\tilde{\beta}_I,\tilde{\beta}_G)
\en
with   $\tilde\beta=-\um\log(\tanh \beta)$.

The phase diagram of this model  was studied long ago
\cite{js} and revisited recently\cite{ggrt}. There is an unconfined 
region surrounded by lines of phase
transitions toward  the Higgs phase and its dual confining phase.
These lines are second order until they are near each other and the
self-dual line, where first order transition occurs. Our simulations are 
of course in the confining phase. 

We measured  two kinds of  observables. The Wilson loop associated to a 
closed path $C$ of links $\ell$ is defined as usual 
\eq
W(C)=\bra\prod_{l\in C}U_{\ell}\ket= \frac1Z\int[\mbox{DU]\,[D}\varphi]
\prod_{l\in C}U_{\ell}\,e^{-S}~~.
\label{vev}
\en
We fitted the numerical data using a refinement of the Ansatz (\ref{Ansatz})
which takes into account the universal contribution of the bosonic string 
quantum fluctuations. Indeed previous numerical work on this subject strongly
suggested \cite{gg} that the nature of the underlying asymptotic string should
be  the same as in the pure gauge model \cite{Caselle:1996ii}. Thus  we put
\eq 
 W(R,T)\simeq c_u\sqrt{\frac{\eta(i)\sqrt{R}}{\eta(\tau)}}
\exp[-2\rho(R+T)-\sigma RT]+c_b\exp[-2\mu (R+T)]~,
\label{refan} 
\en
 where $\eta(\tau)$ is the Dedekind function
\eq 
\eta(\tau)=q^{1/24}\,\prod_{n=1}^{\infty}(1-q^n)~,~  q=e^{2i\pi\tau}~,
\tau=i\frac TR~~.
\en
We also considered the gauge-invariant propagator $G(T)$ of  the static-light
meson. It can be constructed by coupling the product of link
variables along a line of extent $T$ to the matter fields 
$\varphi_0$ and $\varphi_T$ located at both ends  
\eq
G(T)=\bra\varphi_0\prod_{\ell\in T}U_\ell\,\varphi_T\ket~~~.
\label{stali}
\en 
When this line is long enough the asymptotic form of its vacuum 
expectation value is
\eq
 G(T)\simeq c\, e^{-\mu\,T}~,
\label{mu}
\en 
and is well suited to measure $\mu$. We did not apply any kind of smearing 
to the measured operators.

\section{The algorithm  }
\label{algorithm}
The main idea underlying our algorithm relies on the consideration that 
the L\"uscher-Weisz procedure \cite{lw}, reducing the short wavelength 
fluctuations, is capable of an error reduction even if we cannot use their 
argument on the exponential decay of the temporal line correlators, because 
in presence of interacting matter  the confining string breaks. 

We proceed as in the L\"uscher-Weisz algorithm and  split the lattice in 
sub-lattices formed by temporal slices and evaluate Wilson loops via a 
stochastic estimate on these sub-lattices.
To be specific, consider a rectangular $R\times T$ Wilson loop in the
 $(x,t)$ plane. Let $U_0(x,t)$  $(U_1(x,t))$ be the time-like (space-like) 
link variables. The two spatial sides of length $R$ are associated to the
 operators ${\mathbb L}(0)=\prod_{n=1,R/a}{U}_1(n\,a,0) $
and   ${\mathbb L}(T)=\prod_{n=1,R/a}{U}_1(n\,a,T) $ and the operators 
associated to the two temporal sides are conveniently expressed in terms of 
 pairs of time-like links with the same temporal coordinate $t=n\,a$, 
which we call ${\mathbb T}(R,n\,a)=U_0(0,n\,a)U_0(R,n\,a)$.
We adhere to the notation of Ref.\cite{lw}, but we need no matrix 
indices nor complex conjugation, of course. 
The vacuum expectation value defined in Eq.(\ref{vev})
can be rewritten as  
 \eq
W(R,T)=\bra {\mathbb L}(0)\prod_{n=1,T/a}{\mathbb T}(R,n\,a)\,{\mathbb L}
(T)\ket~~.
\label{fette}
\en
Let us split the whole lattice in 
sub-lattices formed by temporal slices of arbitrary thickness.
As observed in ref.\cite{lw},  owing to the locality 
of the action, each slice can be analysed independently of the surrounding 
medium, provided that the link variables of the boundary and the matter field 
configuration are held fixed.
 \par In the present case the expectation value of the operator 
${\mathbb T}(R)$  
in a time-slice formed by $m$ layers is  defined by
\begin{equation}
\begin{split}
[{\mathbb T}_\varphi\{R,j;m\}]\equiv
[{\mathbb T}(R,j\,a) {\mathbb T}(R,(j+1)a)\ldots\\ 
{\mathbb T}(R,(j+m-1)a)]=
\frac1{{\cal Z}_{sub}}
\int[\mbox{DU}]\, e^{-S_{sub}[U,\varphi]}
\end{split}
\end{equation}
where the subscript $sub$ refers to the variables belonging to the sub-lattice.
\par
In our numerical simulations these expectation values are estimated using a 
heat-bath method for the link variables, alternating in a suitable proportion 
with a non-local cluster algorithm for updating the matter field $\varphi$
\cite{gr}.

\par As in pure gauge case one gets identities like
\eq
 [{\mathbb T}_\varphi\{R,j;m\}]\,[ {\mathbb T}_\varphi\{R,(j+m);n\}]=
[{\mathbb T}_\varphi\{R,j;m+n\}]~,
\en
which allow us to rewrite $W(R,T)$ in the form
\eq
W(R,T)=\bra {\mathbb L}(0)\prod_{j=1,T/m\,a}
[{\mathbb T}_\varphi\{R,j;m\}]\; {\mathbb L}(T)\ket~~.
\label{wrt}
\en
where for simplicity it is assumed that $T/a$ is a multiple of $m$. 
In our  simulations we chose $m=2$ and $m=3$. 

\par An updating cycle is organised as follow:
\begin{enumerate}
\item{Generate  field configurations of the whole lattice using the 
heath-bath and the non-local 
cluster updates  as in \cite{gr}};
\item{Hold fixed both  the spatial links along 
some suitably chosen spatial planes and the matter field configuration
and   update  the gauge fields in the interior of the 
slices to estimate $  [{\mathbb T}_\varphi\{R,j;m\}]$;}
\item{Combine the stochastic estimates of the previous step with 
the side operators ${\mathbb L}$ of Eq.(\ref{wrt})} and update  the matter 
field configurations using the non-local cluster algorithm; 
\item Go to 1.
\end{enumerate}
The number $n_w$ of updates of the whole lattice (step 1),  the number 
$n_t$ of updates inside each time-slice (step 2) and the number $n_\varphi$
of updates of the matter field (step 3) are chosen in practice on 
the basis 
of a trial  and error method in the optimisation process 
of the error reduction. For instance, in the  $m=3$  case one updating cycle 
was composed of  $n_t=10^2$ gauge updates for each slice followed by 
one non-local cluster update of the matter configuration (i.e.~$n_\varphi=1$) 
and $n_w=10^5$ gauge updates of the whole lattice. 
The total number cycles generated to extract our 
estimates was $4.0\times 10^4$. All the simulations for this work were 
performed on our cluster employing 20 processors for a total amount of 
7000 hours each.         

\section{Results}
\label{results}
\begin{figure}
\centering
\includegraphics[width=0.95\textwidth]{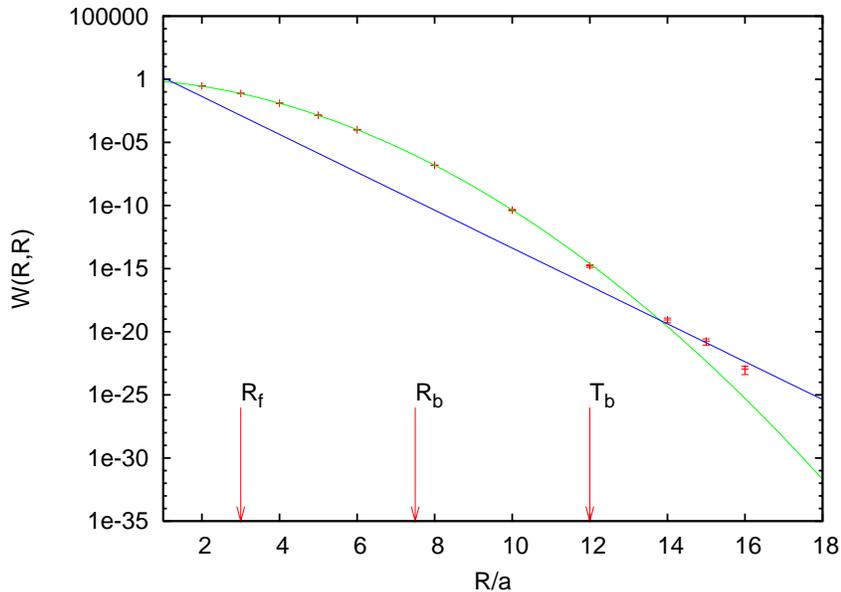} 
%\vspace{5cm}  % amount of vertical space needed
\caption{Square Wilson loop versus  $R$ in a semilogarithmic plot. The 
parabolic line is  the first term of the Ansatz (\ref{refan});  
the straight  line represents the contribution of the broken-string state 
(the second term of the Ansatz) as determined by the fit to 
rectangular Wilson loops at $T=15\,a$ and large $R$.} 
\label{Figure:1}
\end{figure}

We first explored a wide region of the confining phase in order to see whether 
there exists a parameter range  where the lower bound (\ref{survival}) is 
accessible to realistic simulations.  For this purpose we 
measured the vacuum expectation value of square Wilson loops 
$ W(T,T)$ and the 
static-light meson correlators $ G(T)$ in a wide range of $T$ 
on a $40^3$ cubic lattice in order to estimate the string tension $\sigma$ 
and the string breaking scale $R_b$ trough Eq.(\ref{rb}).

The square Wilson loop data were fitted to the first term of Eq.(\ref{refan}) 
by progressively eliminating the data of lower $T$ until stable parameters 
were obtained. While doing this, the scale $R_f$ of string formation was 
determined by picking out the value of $T=R_f$ such that the exclusion 
(inclusion) of $\bra W(R_f,R_f)\ket$ in the fit resulted in a good (bad) 
$\chi^2$ test. In all the cases considered there was no ambiguity in the 
choice of $R_f$, being the  corresponding variation of 
$\chi^2$  large enough.         
\par
In this way we selected the point $\beta_G=0.650$, $\beta_I=0.235$ 
corresponding, in lattice spacing units $a$, to
\eq 
\sigma a^2=0.2054(21),~\mu a=0.8637(10),~ 
R_f\simeq 3 a ,~ R_b\simeq 7.5 a,~c_u=1.19(5)~. 
\label{data}
\en
Here we concentrated our computational efforts enlarging the lattice size to 
$50^3$ and measuring systematically all the 
rectangular Wilson loops   with $4a\leq R\,\leq 18a$ and $T/a$ a multiple of
 2 or 3.
\par
\begin{figure}
\centering
\includegraphics[width=0.95\textwidth]{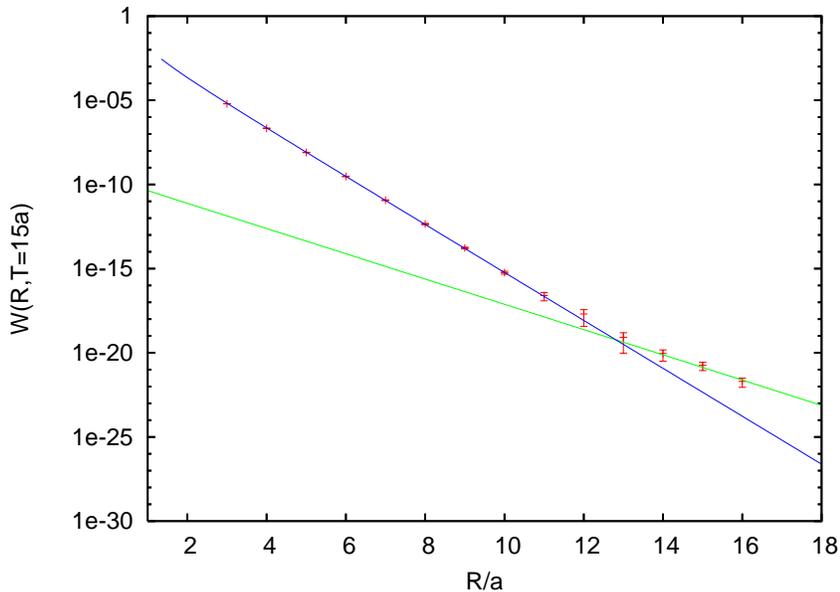} 
\caption{Wilson loop data versus $R$, for $T=15\,a$. The steepest line is not a
 fit, but represents the first term of Eq.(\ref{refan}) as determined by  
fitting the square Wilson loops. The slope of the other line is twice the 
mass of the static-light meson. The only free parameter is the intercept of 
this line.}
\label{Figure:2}
\end{figure}
\begin{figure}
\centering
\includegraphics[width=0.95\textwidth]{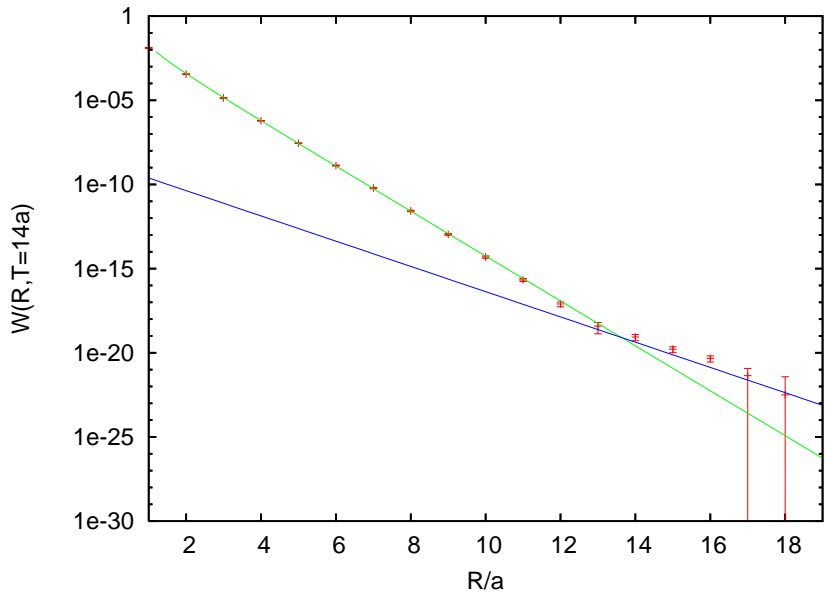} 
%\vspace{5cm}  % amount of vertical space needed
\caption{Wilson loop data versus $R$ for $T=14\,a$. We did not fit any 
parameters but used those of the previous figure combined with the Ansatz 
(\ref{refan}).}
\label{Figure:3}
\end{figure}

In Fig.\ref{Figure:2} and Fig.\ref{Figure:3} we show $ W(R,T)$ at 
$T=15\,a$ and $T=14\,a$ as a function of $R$. Notice the abrupt change in 
slope, signalling string breaking. The steepest line is not a best fit, but is 
drawn  using the first term of Eq.(\ref{refan}) (unbroken-string term) where 
the parameters are those fitting the square Wilson loops. The slope of 
the other line is given by $2\mu$. The only
parameter used to fit the data of Fig.\ref{Figure:2} is $c_b$  of 
Eq.(\ref{refan}). We estimated $c_b=43(5)$. This value  is 
used in turn to account for the  data of Fig.\ref{Figure:3}. Inserting these values in Eq.(\ref{bound}) we see that the upper bound is nearly saturated and 
the string-breaking scale in Fig.\ref{Figure:1} is correspondingly slightly 
bigger than $T_b$. 

\section{Conclusion}
The string breaking phenomenon in gauge theories coupled with matter is 
hardly detectable while measuring Wilson loops only. This reflects the 
poor overlap of 
the Wilson loop operator with the broken-string state. A simple way to 
represent such a behaviour is to
assume that the vacuum expectation value of a rectangular Wilson loop 
$W(R,T)$  is  the sum of two terms (see Eq.(\ref{Ansatz})), 
one decaying with an area law which prevails at intermediate scales and the 
other  obeying  a perimeter law which describes the expected asymptotic 
behaviour. It is worth noting that the logarithm of the former term is  a 
hyperbola in the $R,T$ plane, while the log of the latter is a straight line; 
they intersect at two points. One intersection represents the cross over 
to the broken-string state. The other intersection cannot be 
completely arbitrary: in order not to spoil the area law at intermediate 
distances one is forced to put an upper bound of the overlap to the 
broken-string state of the Wilson operator. Such an upper bound demonstrates
{\sl a posteriori} why it is so difficult to see string breaking in current 
simulations using Wilson loops only: even if the most favourable 
conditions were met, namely the saturation of our upper bound 
and the optimisation of the signal  by measuring square Wilson loops, 
the minimal distance at which string-breaking is visible is $T_b=2R_b-R_f$ 
(see Eq.(\ref{tb})),  where $R_b$ is the string-breaking scale and $R_f$ 
is the minimal scale of string formation. From a computational point of view 
$T_b$ is too large for the current updating algorithms. This explains why 
earlier studies on Wilson loops in gauge theories coupled to matter, 
which did not used a variance reduction method, failed to observe 
string-breaking.

Adapting  the L\"uscher-Weisz variance reduction method to the 
3-D $\Z_2$ gauge Higgs model, which is perhaps the simplest gauge theory 
coupled to matter, we found a clean and  beautiful signal of 
string breaking in the Wilson operator in a region where our upper bound 
is nearly saturated.

\end{document}